\documentclass[twocolumn,showpacs,preprintnumbers,amsmath,amssymb]{revtex4}
\usepackage{dcolumn}% Align table columns on decimal point
\usepackage{bm}% bold math
\usepackage{graphicx}% Include figure files
\DeclareGraphicsExtensions{.jpg,.pdf,.mps,.png,.eps,.ps,.EPS}
\def\mean#1{\left\langle #1\right\rangle}

\begin{document}

\title{Coarse-Graining and Self-Dissimilarity of Complex Networks}

\author{Shalev Itzkovitz, Reuven Levitt, Nadav Kashtan, Ron Milo, Michael Itzkovitz, Uri Alon}
\affiliation{Departments of Molecular Cell Biology and Physics of
Complex Systems, Weizmann Institute of Science, Rehovot, Israel 76100\\
}

\begin{abstract}
Can complex engineered and biological networks be coarse-grained
into smaller and more understandable versions in which each node
represents an entire pattern in the original network? To address
this, we define coarse-graining units (CGU) as connectivity patterns
which can serve as the nodes of a coarse-grained network, and
present algorithms to detect them. We use this approach to
systematically reverse-engineer electronic circuits, forming
understandable high-level maps from incomprehensible transistor
wiring: first, a coarse-grained version in which each node is a gate
made of several transistors is established. Then, the coarse-grained
network is itself coarse-grained, resulting in a high-level
blueprint in which each node is a circuit-module made of multiple
gates. We apply our approach also to a mammalian protein-signaling
network, to find a simplified coarse-grained network with three main
signaling channels that correspond to cross-interacting MAP-kinase
cascades. We find that both biological and electronic networks are
'self-dissimilar', with different network motifs found at each
level. The present approach can be used to simplify a wide variety
of directed and nondirected, natural and designed networks.

\end{abstract}

\maketitle

\section{Introduction}
In both engineering and biology it is of interest to understand the
design of complex networks~\cite{Strogatz 2001,Albert,newman_siam},
a task known as 'reverse-engineering'. In electronics, digital
circuits are top-down-engineered starting from functional blocks,
which are implemented using logic gates, which in turn are
implemented using transistors ~\cite{Horowitz}. Reverse-engineering
of an electronic circuit means starting with a transistor map and
inferring the gate and block levels. Current approaches to
reverse-engineering of electronic circuits usually require prior
knowledge of the library of modules used for forward-engineering
~\cite{Kundu,Hansen}.\\
In biology, increasing amounts of interaction networks are being
experimentally characterized, yet there are few approaches to
simplify them into understandable blueprints
\cite{McAdams,Hartwell,D'Haeseleer,Kohn,Ravasz,Csete,Rives,tyson,Gardner,Beer,Kalir,Alon}.\\
Here we present an approach for simplifying networks by creating
coarse-grained networks in which each node is a pattern in the
original network. This approach is based on network motifs,
significant patterns of connections that recur throughout the
network ~\cite{Shen-orr,Milo,Milo 2004,Kashtan}. We define
coarse-graining units, CGUs, which can be used as nodes in a
coarse-grained version of the network. We demonstrate this approach
by coarse-graining an electronic and a biological network.

\textbf{Definition of CGUs}: CGUs are patterns which can optimally
serve as nodes in a coarse-grained network. One can think of CGUs as
elementary circuit components with defined input and output ports,
and internal computational nodes. The set of CGUs comprise a
dictionary of elements from which a coarse-grained version of the
original network is built.\\
The coarse-grained version of the network is a new network with
fewer elements, in which some of the nodes are replaced by CGUs. Our
approach to define CGUs is loosely analogous to coding principles
and to dictionary text compression techniques~\cite{Barron,Bell}.
The goal is to choose a set of CGUs that (a) is as small as
possible, (b) each of which is as simple as possible, and which (c)
make the coarse-grained network as small as possible. These three
properties can be termed 'conciseness', 'simplicity' and 'coverage'.
Conciseness is defined by the number of total CGU types in the
dictionary set. Coverage is the number of nodes and edges eliminated
by coarse-graining the network using the CGUs. To define simplicity,
we describe each occurrence of the subgraph, $G$, as a 'black box'.
The black box has input ports and output ports, which represent the
connections of $G$ to the rest of the network, $R$ (Fig 1).
\begin{figure}[!hbp]
\begin{center}
\includegraphics[width = 70 mm, height = 80 mm ]{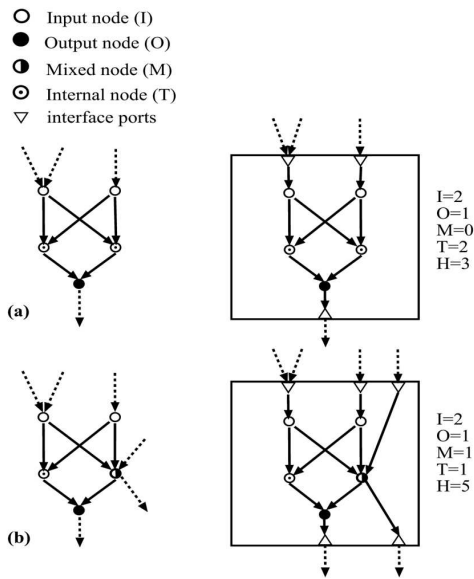}
%\resizebox{8cm}{!}{\includegraphics{Fig1.eps}}
\end{center}
\caption{Black box representation of a subgraph and the classes of
nodes and ports. The nodes of the subgraph (numbered 1-5) are
classified into input (I), output (O), internal (T) and mixed (M)
nodes according to the edges that connect them to the rest of the
network (dashed arrows). The subgraph is represented as a black
box with input and output ports (right side of figure). The
complexity-measure $H$ is the total number of ports. \textbf{a.}
Subgraph with no mixed nodes. The connectivity profile vector is
(I,I,T,T,O) \textbf{b.} subgraph with a mixed node. The
connectivity profile vector is (I,I,T,M,O).} \label{fig1}
\end{figure}
There can be four types of nodes in $G$ : input nodes that receive
only incoming edges from $R$, output nodes that have only outgoing
edges to $R$, internal nodes with no connection to $R$, and mixed
nodes with both incoming and outgoing edges to $R$. To obtain a
minimal loss of information, a coarse-grained version of $G$
includes ports, which carry out the interface to the rest of the
network. The number of ports in the black box representing $G$ is:
\begin{eqnarray}
H=I+O+2M\label{eq1}
\end{eqnarray}
where $I$ is the number of input nodes, $O$ the number of output
nodes and $M$ the number of mixed nodes (internal nodes do not
contribute ports and each mixed node contributes two ports). The
lower the number of ports, $H$, the more 'simple' the CGU.\\
After defining simplicity, coverage and conciseness, one can choose
the optimal set of CGUs. To choose the optimal set of CGUs, we
maximize a scoring function which is defined for a set of $N$ CGUs:
\begin{eqnarray}
 S=E_{covered} + \alpha \Delta{P} - \beta N - \gamma \sum_{i=1}^N{T_i}\label{eq2}
\end{eqnarray}
$E_{covered}$ is the number of edges covered by all occurrences of
the CGUs, and therefore eliminated in the coarse-grained network.
$N$ is the number of distinct CGUs, $T_i$ is the number of
internal nodes in the $i$-th CGU. $\Delta{P}$ is the difference
between the number of nodes in the original network and the number
of nodes and ports in the coarse-grained network:
\begin{eqnarray}
\Delta{P}=P_{covered}-\sum_{i=1}^N{n_i H_i}\label{eq3}
\end{eqnarray}
where $P_{covered}$ is the number of nodes covered by all
occurrences of the CGUs, $n_i$ is the number of occurrences in the
network of CGU $i$, and $H_i$ is the number of ports of CGU $i$.
Using this we obtain:
\begin{eqnarray}
 S=[ E_{covered} + \alpha P_{covered} ]-[ \alpha \sum_{i=1}^N{n_i H_i}+\beta N + \gamma \sum_{i=1}^N{T_i} ]\label{eq4}
\end{eqnarray}
The scoring function has two terms: The first term, corresponding to
coverage, corresponds to the simplification gained by
coarse-graining, while the second term, corresponding to simplicity
and conciseness, quantifies the complexity of the CGU dictionary.
Maximizing $S$ favors use of a small set of CGUs, preferentially
those that appear often, with many internal nodes and few mixed
nodes (since internal nodes do not contribute ports to $H_i$, and
mixed nodes contribute two ports).\\
The last term in the scoring function, which is the total number of
internal nodes in the dictionary, bounds the CGU size and prevents
the trivial solution where the entire network is replaced by a
single complex node. $\alpha$,$\beta$,$\gamma$ are parameters that
can be set for various degrees of coarse-graining (The results below
are insensitive to varying these parameters over 3 orders
of magnitude).\\
We use a simulated annealing approach~\cite{Kirkpatrick} to find the
optimal set of CGUs for coarse-graining : There is potentially a
huge number of subgraphs that can serve as candidate CGUs. To reduce
the number of candidate subgraphs, and to focus on those likely to
play functional roles, we consider only subgraphs that occur in the
network significantly more often than in randomized networks:
network motifs~\cite{Shen-orr,Milo,Milo 2004,Kashtan}. A candidate
set of CGUs is obtained by first detecting all network motifs of
$3-6$ nodes (Appendix A). The nodes of every occurrence of each
motif are classified to one of the 4 types
(input/output/internal/mixed). This defines a connectivity profile
for each occurrence. For example, the two subgraphs in Fig 1 have
the profiles (I,I,T,T,O) and (I,I,T,M,O),
where I,T,O,M represent input, internal, output and mixed nodes respectively.\\
The occurrences of each motif are then grouped together according to
their profile to form a CGU candidate \cite{footnote0}. A CGU
candidate of $n$ nodes is thus characterized by its topology (an
$n*n$ adjacency matrix) and by a $n$-length profile vector of node
classifications (Fig 3).\\
In the simulated annealing optimization algorithm, each CGU
candidate is assigned a random spin variable which is either 1 if
all its occurrences participate in the coarse-graining or 0
otherwise. CGU candidates with spin 1 compose the "active set". At
each step a spin is randomly chosen and flipped, and the
coarse-graining score for the new active set is computed. The active
set is updated according to a Metropolis Monte-Carlo procedure
~\cite{newman,footnote1}.\\
Once an optimal set of CGUs is found a coarse-grained
representation of the original network is formed by replacing each
occurrence of a CGU with a node (Appendix B). Generally the
coarse-grained representation is a hybrid in which some nodes
represent CGUs, and other nodes are the original nodes. The
algorithm can be repeated on the coarse-grained representation, to
obtain higher levels of
coarse-graining.\\
\begin{figure}[!hbp]
\begin{center}
\includegraphics[width =90 mm, height = 70 mm ]{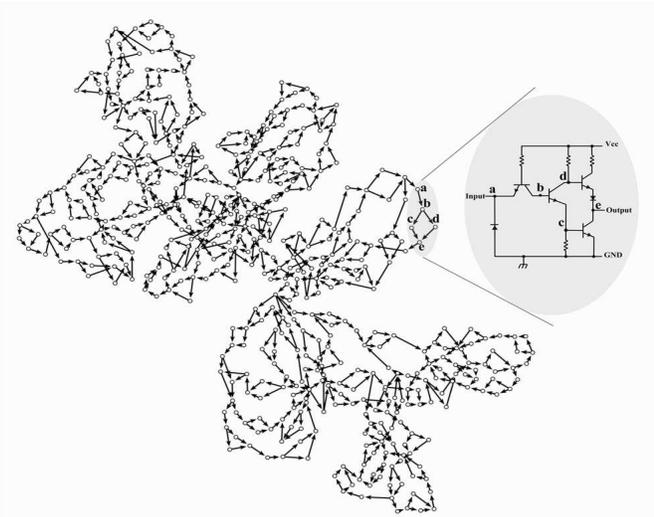}
\end{center}
\caption{Transistor level map of an 8-bit binary counter, (ISCAS89
circuit S208 \cite{Brglz}). Nodes are junctions between
transistors, and directed edges represent wire connections.
Highlighted is a subgraph that represents the transistors that
make up one NOT gate.}\label{fig2}
\end{figure}
\begin{figure}[!hbp]
\begin{center}
\includegraphics[width = 70 mm, height = 70 mm ]{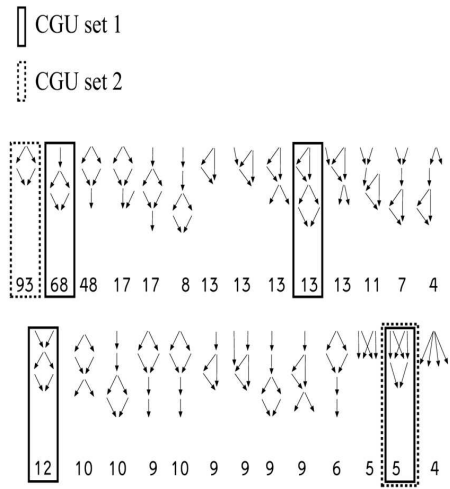}
%\resizebox{8cm}{!}{\includegraphics{Fig1.eps}}
\end{center}
\caption{A partial set of the network motif candidate CGUs for the
transistor level network. The number of occurrences of each motif in
the transistor network is shown. The optimal CGU dictionary consists
of 4 units (solid boxes - CGU set $1$ $\alpha=0.2, \beta=20,
\gamma=0.01$). A second optimal solution consisting of 2 units,
which is found for high values of $\beta$ is also shown (dashed
boxes - CGU set 2 $\alpha=0.2, \beta=500, \gamma=0.01$). Note that
several CGU candidates share the same motif topology. They differ by
their connectivity profile vectors (input/output/internal/mixed)}
\label{fig3}
\end{figure}
\begin{figure}[!hbp]
\begin{center}
\includegraphics[width = 70 mm, height = 70 mm ]{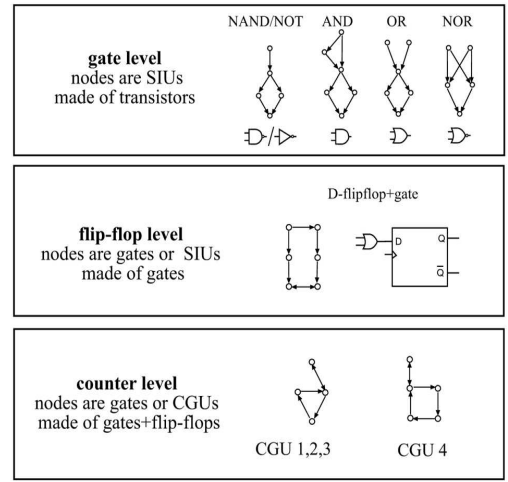}
%\resizebox{8cm}{!}{\includegraphics{Fig1.eps}}
\end{center}
\caption{The CGUs found in the different coarse-grained levels of
the electronic circuit. At the gate level the CGUs are the TTL
implementation of AND, OR, NAND, NOR and NOT gates (NAND and NOT
differ by the type of transistor at the input). At the flip-flop
level, a single CGU, occurring 8 times is found. This CGU
corresponds to the 5-gate implementation of a D-flip-flop with an
additional gate at the input. At the counter level, two CGU
topologies are found: Seven occurrences of a 3-node feedback
loop+mutual edge, and one occurrence of a 4-node feedback
loop+mutual edge, representing CGU4.} \label{fig4}
\end{figure}
\begin{figure*}[]
\begin{center}
\includegraphics[width = 150 mm, height = 100 mm ]{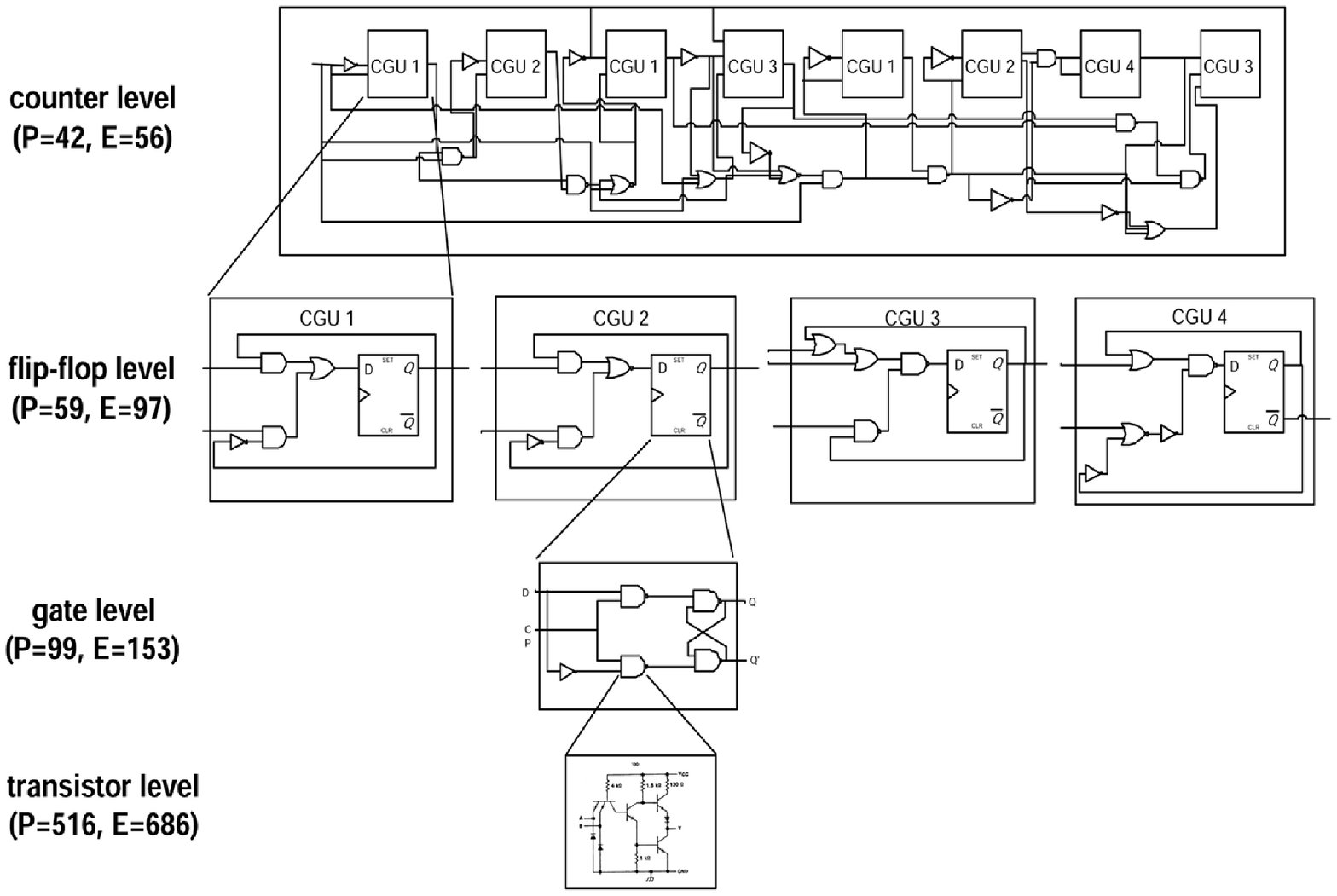}
\end{center}
\caption{Four levels of representation of the electronic circuit.
In the transistor level, nodes represent transistor junctions. In
the gate level, nodes are CGUs made of transistors, each
representing a logic gate. Shown is the CGU that corresponds to a
NAND gate. In the flip-flop level, nodes are either gates or a CGU
made of gates that corresponds to a D-type flip-flop with an
additional logic gate at its input. In the counter level, each
node is a gate or a CGU of gates/flip-flops that corresponds to a
counter subunit. Numbers of nodes ($P$) and edges ($E$) at each
level are shown.}\label{fig5}
\end{figure*}
Note that the coarse-graining problem is quite different from the
well-studied circuit partitioning problem \cite{Alpert}, and from
the detection of community structure in networks
~\cite{Girvan,Zhou2,Zhou}. These algorithms seek to divide networks
into subgraphs with minimal interconnections, usually resulting in a
set of distinct and rather complex subgraphs. In contrast,
coarse-graining seeks a small dictionary of simple subgraph types in
order to help understand the function of the network in terms of
recurring independent building blocks. An analogy is the detection
of words in a text, from which spaces and punctuation marks have
been removed, without prior knowledge of the language.

\section{Coarse-graining of an electronic circuit}
To demonstrate the coarse-graining approach we analyzed an
electronic circuit derived from the ISCAS89 benchmark circuit
set~\cite{Brglz,Cancho}. The circuit is a module used in a digital
fractional multiplier ($S208$~\cite{Brglz}). The circuit is given as
a netlist of 5 gate types (AND, OR, NAND, NOR, NOT) and a
D-flipflop(DFF). To synthesize a transistor level implementation of
this circuit (Fig 2) we first replaced every DFF occurrence with a
standard implementation using 4 NAND gates and one NOT
gate~\cite{Horowitz}. All gates were then replaced with their
standard transistor-transistor logic (TTL)
implementation~\cite{texas}, where nodes represent junctions between
transistors (for this purpose resistors and diodes were ignored, as
were ground and
Vcc). The resulting transistor network (Fig 2) has 516 nodes and 686 edges.\\
Four CGUs were detected in the transistor level, each with five or
six nodes (Fig 3,4a). These patterns correspond to the transistor
implementations of the five basic logic gates AND, NAND, NOR, OR
and NOT (Fig 4a). These CGUs were used to form a coarse-grained
version of the network in which each node is a CGU. In this case
coverage was complete, and all of the original
nodes were included within CGUs. This network, termed 'gate-level network' had 99 nodes and 153 edges.\\
We next iterated the coarse-graining process, by applying the
algorithm to the gate-level network. One CGU with six nodes
(gates) was detected. This CGU corresponds to a D-flip-flop with
an additional logic gate (Fig 4b). A 'flip-flop level'
coarse-grained network was then formed with nodes which were
either gates or flip-flops. This
network had 59 nodes and 97 edges.\\
We applied the coarse-graining algorithm again to the flip-flop
level network.Two types of CGUs were found (Fig 4c), which
correspond to units of a digital counter. Using these CGUs, we
constructed the highest-level coarse-grained network in which each
node is either a CGU or a gate. This network, depicted in Fig 5 top
panel, had 42 nodes and 56 edges. Thus, the highest-level
coarse-grained network has 12-fold fewer nodes and edges than the
original transistor-level network. This high-level map corresponds
to sequential connections of binary counter units, each of which
halves the frequency of the binary stream obtained from the previous
unit. This map thus describes an 8-bit counter~\cite{Nagle}.\\
In other electronic circuits, we find other CGUs, including a XOR
built of 4 NAND gates \cite{Horowitz},\cite{Kashtan} (data not
shown). The coarse-graining approach appears to automatically detect
favorite modules used by electronic engineers.

\section{Coarse-graining of biological networks}
Recent studies have shown that biological networks contain
significant network motifs~\cite{Shen-orr,Milo,Milo 2004,Kashtan}.
Theoretical and experimental studies have demonstrated that each
network motif performs a key information processing
function~\cite{Kalir,Alon,Shen-orr,Mangan,Ronen,Mangan2,Zaslaver,Lahav,rosenfeld}.
A coarse-grained version of biological networks is of interest
because it would provide a simplified representation, focused on
these important sub-circuits. However, whereas electronic circuits
are composed of exact copies of library units, in biology the
recurring black boxes may not be of precisely the same structure. In
addition, the characterization of signaling and regulatory networks
is currently incomplete due to experimental limitations. Thus a more
flexible definition of CGUs is needed \cite{berg}. To address these
issues we modify our algorithm by allowing each CGU to represent a
family of subgraphs, which share a common architectural theme. Thus,
the CGUs are \emph{probabilistically generalized network motifs
(PGNM)}: network motifs of different sizes which approximately share
a
common connectivity pattern.\\

\textbf{Probabilistic generalization of network motifs}: To define
PGNMs, we must first discuss the concept of block-models
\cite{white}-\cite{panning}. A block-model is a compact
representation of a subgraph. It consists of two elements : 1) a
partition of the subgraph nodes into discrete subsets, called
\emph{roles} \cite{Kashtan}. 2) a statement about the presence or
absence of a connection between roles (Fig 6). A subgraph of $n$
nodes can be described by an adjacency matrix $G$, where $G_{ij}=1$
if a directed edge exists from node $i$ to node $j$, and $G_{ij}=0$,
if there is no connection. A block-model partitions the $n$ nodes
into $m\leq{n}$ roles according to \emph{structural equivalence}.
Two nodes are structurally equivalent if they share exactly the same
connections to all other nodes. The block model is an $m*m$ matrix
$A$, where $A_{IJ}=1$ means that all nodes which share role $I$ have
a directed connection to all nodes
which share role $J$ (Fig 6).\\
In large subgraphs of real-world networks, perfect structural
equivalence is not always seen. A block-model can still be used as
an idealized structure which can be compared to a given subgraph.
The distance between a subgraph and a proposed block-model, can be
defined as \cite{panning}:
\begin{eqnarray}
d=\frac{S_{W}}{S_T} \label{eq5}
\end{eqnarray}
where $S_{W}$ is the within-block sum of squares :
\begin{eqnarray}
%WBSS=\sum_I{\sum_J{\sum_{i\in{I},j\in{J}}{(G_{ij}-\mean{G_{IJ}})^2}}}
S_{W}=\sum_I{\sum_J{\sum_{i\in{I},j\in{J}}{(G_{ij}-\mean{G_{IJ}})^2}}}
\label{eq6}
\end{eqnarray}
$\mean{G_{IJ}}$ is the mean of the adjacency matrix values in
block $\{I,J\}$, and $S_{T}$ is the total sum of squares :
\begin{eqnarray}
S_{T}=\sum_{i,j}{(G_{ij}-\mean{G})^2} \label{eq7}
\end{eqnarray}
where $\mean{G}=\sum{G_{ij}}/n^2$ is the mean value of
$G$\cite{footnote2}. A subgraph with $d=0$ is perfectly described by
its block model.
\begin{figure}[!hbp]
\begin{center}
\includegraphics[width = 80 mm, height = 100 mm ]{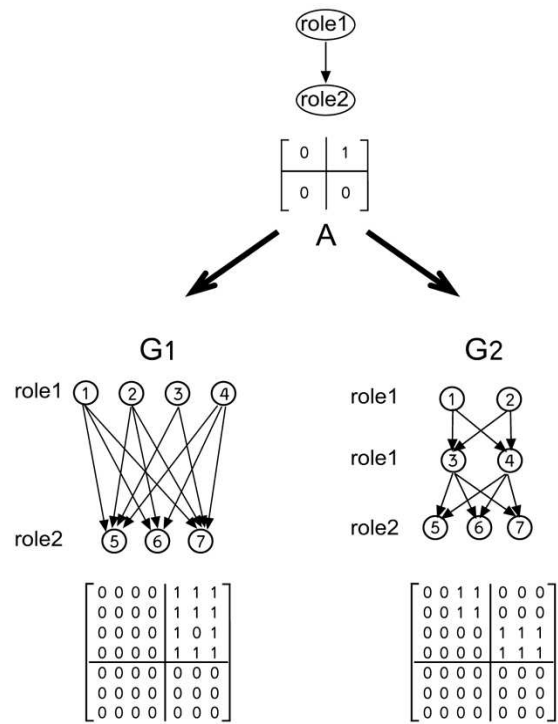}
\end{center}
\caption{A block-model (top) and two subgraphs, one which fits the
block-model ($G1$, bottom left) and one which does not ($G2$, bottom
right). $G1$ has $7$ nodes and $2$ roles (nodes $1-4$ share role $1$
and nodes $5-7$ share role $2$). Its adjacency matrix is shown
below, with lines indicating the block-model partition. An edge
between node 3 and node 6 is missing for a perfect fit to the
proposed block-model. The distance between the block matrix and the
adjacency matrix is $d=0.1075$. The right subgraph, $G2$ does not
fit the proposed block-model $A$. The distance between the block
matrix and the adjacency matrix is $d=0.7538$. An alternative
block-model with 3 roles - $(\{1,2\},\{3,4\},\{5,6,7\})$ would
perfectly fit this subgraph, with $d=0$ . Both of these subgraphs
are aggregates of a $4$-node bifan subgraph (Fig 7).}\label{fig6}
\end{figure}
\begin{figure}[!hbp]
\begin{center}
\includegraphics[width = 70 mm, height = 80 mm ]{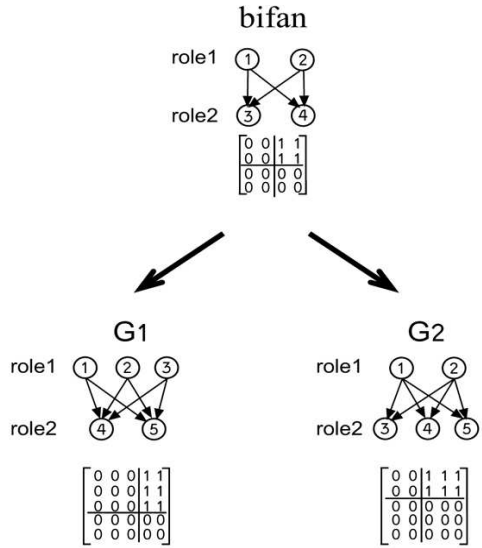}
\end{center}
\caption{Topological generalizations of the bifan \cite{Shen-orr}
subgraph and their adjacency matrices. The bifan subgraph has two
roles - nodes $1,2$ share role $1$ and nodes $3,4$ share role $2$.
Lines indicate the block-model partition. Below are two generalized
subgraphs obtained by role replication \cite{Kashtan}. Subgraph $G1$
(left) is obtained by replicating the first role, with its
connections. Subgraph $G2$ (right) is obtained by replicating the
second role, with its connections. Adjacency matrices and
block-model partitions are shown. The role-replication operation
extends a subgraph while keeping a perfect fit to its
block-model.}\label{fig6new}
\end{figure}
For example, subgraph $G1$ in Fig $6$ has $n=7$ nodes. It can be
described by a block-model with $m=2$ roles. Nodes $1-4$ are
assigned the first role and nodes $5-7$ are assigned the second
role. The distance between the subgraph and the proposed block-model
is $d=0.1075$. Fig $6$ also
shows a subgraph, $G2$, which is far from the proposed block model ($d=0.7538$).\\
Finding the best block-model to fit arbitrary connectivity data is a
combinatorially complex problem~\cite{white,wasserman,panning},
requiring exhaustive testing of different assignments of nodes to
roles. However, an efficient algorithm to detect PGNMs can be formed
based on the fact that small network motifs in biological networks
aggregate to form \emph{network motif topological
generalizations}~\cite{Kashtan,dobrin}. Topological generalizations
are subgraphs obtained from smaller network motifs, by replicating
one or more of their roles, together with its connections
~\cite{Kashtan} (Fig $7$). An algorithm to detect PGNMs is described
in Appendix C.

\begin{figure}[!hbp]
\begin{center}
\includegraphics[width = 70 mm, height = 60 mm ]{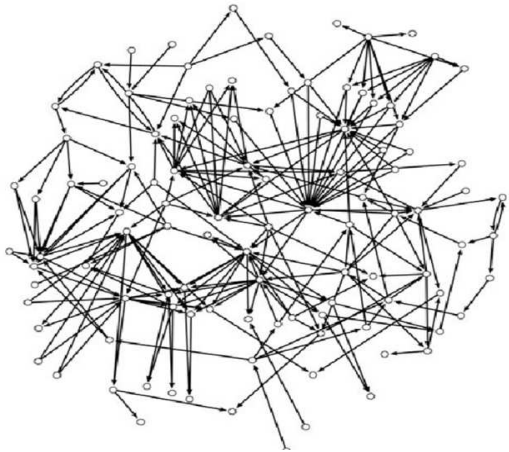}
\end{center}
\caption{A network of signal-transduction pathways in mammalian
cells.}\label{fig7}
\end{figure}

\begin{figure*}[]
\begin{center}
\includegraphics[width = 160 mm, height = 110 mm ]{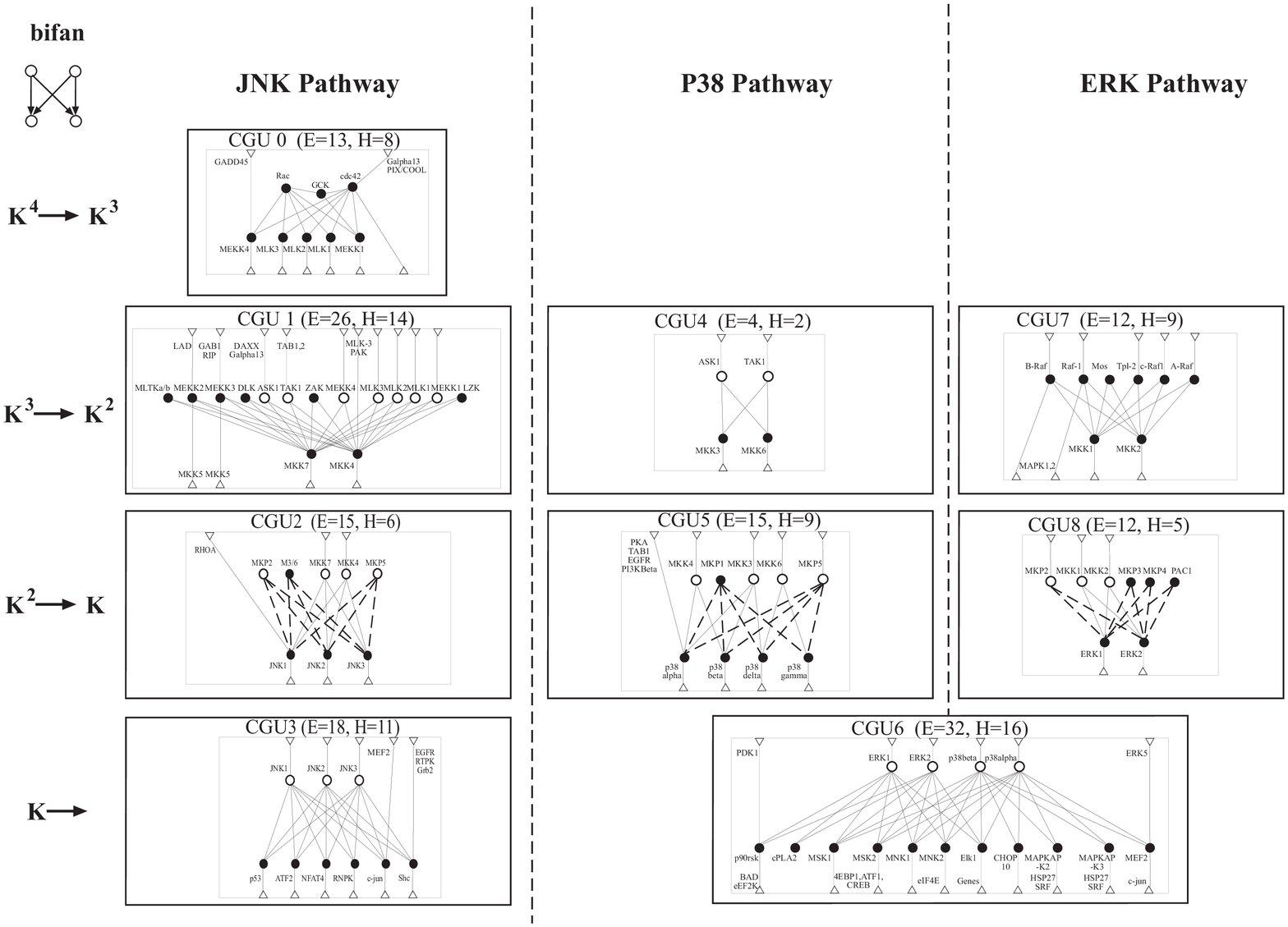}
\end{center}
\caption{CGUs in the signal-transduction network. One CGU is found,
the 4-node bifan with 9 PGNMs, numbered CGU0-CGU8. Solid arrows
represent positive (kinase) interactions, dashed arrows represent
negative interactions (phosphatase). Empty circles represent
duplicated nodes (nodes which participate in more than one PGNM).
$K$, $K^2$, $K^3$ and $K^4$ represent MAP-kinase, kinase-kinase,
kinase-kinase-kinase etc.}\label{fig9}
\end{figure*}

\begin{figure*}[]
\begin{center}
\includegraphics[width = 100 mm, height = 70 mm ]{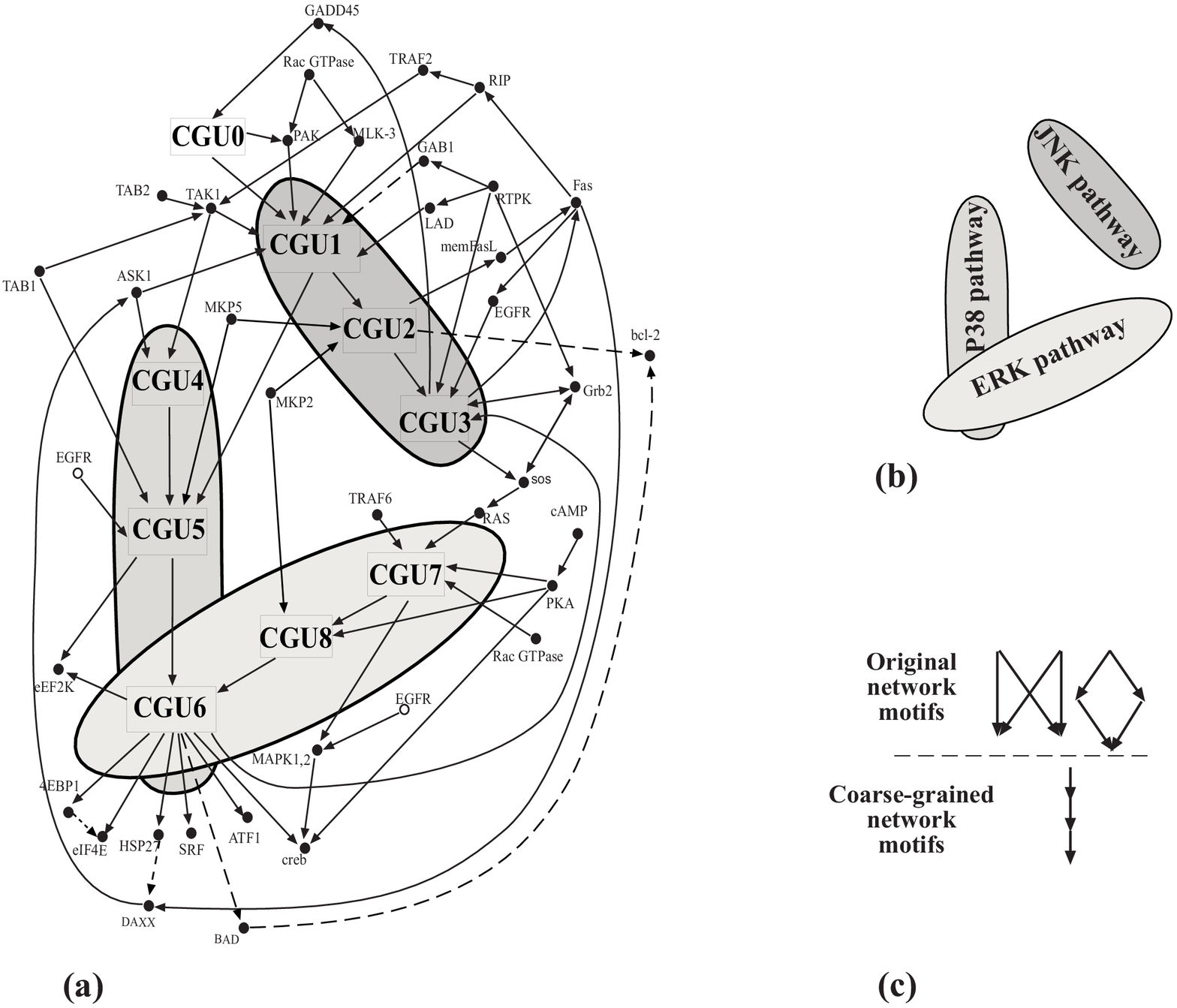}
\end{center}
\caption{\textbf{a.} Coarse-grained version of the
signal-transduction network. Three signaling channels made of
cascades of the CGU occurrences are highlighted. Solid arrows
represent positive (kinase) interactions, dashed arrows represent
negative interactions (phosphatase). EGFR and PKA have been drawn
more than once for clarity. \textbf{b.} The three signaling
channels. \textbf{c.} The network motifs~\cite{Milo} found at the
two levels.}\label{fig10}
\end{figure*}

To determine the optimal dictionary of CGUs, including the PGNMs,
we use the following modified version of the scoring function of
Eq 2 :
\begin{eqnarray}
 S=E_{covered} + \alpha \Delta{P} - \beta N - \gamma \sum_{i=1}^N{T_i} - \delta \sum_{i\in\{CGU_{g}\}}^N{d_i}\label{eq8}
\end{eqnarray}
$N$, the number of CGUs, is the number of basic motifs used.
$CGU_{g}$ includes the set of all PGNMs based on the CGUs. Each CGU
can give rise to several PGNMs of different sizes~\cite{footnote3}.

\section{CGUs in a protein-signaling network}
We analyzed a database of mammalian signal transduction pathways
~\cite{Huang,Bhalla,Charette,Levine,Pearson,nishida,wang,Kyriakis,stke}
based on the Signal Transduction Knowledge Environment \cite{stke}.
This dataset contains 94 proteins and 209 directed interactions (Fig
8). The optimal coarse graining is based on a single motif - the
$4$-node bifan (Fig 9). Thus $N=1$. We find 9 occurrences of PGNMs
based on the bifan, labelled CGU0-CGU8 which share a common design
consisting of a row of input nodes with overlapping interactions to
a row of output nodes (Fig 9). The input and output rows in these
CGUs sometimes represent proteins from the same sub-family (eg.
JNK1,JNK2 and JNK3 in CGU 3), and in other cases they represent
proteins from different sub-families (ERK and p38 in CGU 6). This
type of structure allows hard-wired combinatorial activation and
inhibition of outputs. Similar structures were described in
transcription regulation networks ('dense overlapping regulons'
\cite{Shen-orr}).\\
Using this CGU, the signaling network can be coarse-grained (Fig
10a), showing three major signaling channels (Fig 10b). These
channels correspond to the well-studied ERK, JNK and p38 MAP-kinase
cascades, which respond to stress signals and growth factors~\cite{Huang,Bhalla,Charette,Levine,Pearson,nishida,wang,Kyriakis}.\\
Each channel is made of three CGUs in a cascade. In each cascade,
the top and bottom CGUs contain only positive (kinase) interactions,
and the middle CGU contains both positive and negative (phosphatase)
interactions. The p38 and ERK channels intersect at CGU 6. The MAPK
phosphatase 2 (MKP2) participates in both the JNK pathway (CGU2) and
the ERK pathway (CGU8), whereas MAPK phosphatase 5 (MKP5)
participates in both JNK pathway (CGU2) and the P38 pathway (CGU5).
The MAPKKK ASK1 and TAK1 are shared by both the JNK pathway (CGU1)
and P38
pathway (CGU4)~\cite{nishida,wang}.\\
\section{Self-Dissimilarity of network structure} Interestingly,
the coarse-grained signaling network displays a different set of
network motifs than the original network, with prominent cascades
(Fig 10c). Similarly, the electronic network displayed different
CGUs at each level (Fig 4). These networks are therefore
\emph{self-dissimilar}~\cite{Wolpert,Carlson}: the structure at each
level of
resolution is different.\\

\section{Discussion} We presented an approach for coarse-graining
networks in which a complex network can be represented by a
compact and more understandable version. Performing an
optimization on the space of network motifs of different sizes, we
found optimal units for coarse-graining, CGUs, which allow a
maximal reduction of the network, while keeping a concise and
simple dictionary of elements. We demonstrated that this method
can be used to fully reverse-engineer electronic circuits, from
the transistor level to the highest module level, without prior
knowledge of the library components used to create them.\\
For biological networks, where modularity may be less stringent than
electronic circuits, we modified the algorithm to seek a
coarse-grained network, using a small set of structures of different
sizes that form probabilistically generalized network motifs. Using
this approach, a coarse-grained version of a mammalian signaling
network was established, using one CGU composed of cross-activating
MAP-kinases. In the coarse-grained network one can easily visualize
intersecting signalling pathways and feedback loops. The present
approach allows a simplified coarse-grained view of this important
signaling network, showing the major signaling channels, and
specifies the recurring circuit element (CGU) that may characterize
protein signaling pathways in other
cellular systems and organisms.\\
Biological and electronic networks are both
self-dissimilar~\cite{Wolpert,Carlson}, showing different network
motifs on different levels. This contrasts with views based on
statistical-mechanics near phase-transition points which emphasize
self-similarity of complex
systems.\\
It is important to stress that not every network can be effectively
coarse-grained, only networks with particular modularity and
topology. The method can readily be applied to nondirected networks.
It would be interesting to apply this approach to additional
biological networks, to study the systems-level function of each
CGU, and to study which networks evolve to have a topology that can
be coarse-grained. \acknowledgements We thank J. Doyle, H. McAdams,
J. E. Ferrell, Y. Srebro, E. Dekel, and all members of our lab for
valuable discussions. We thank the Israel Science Foundation, NIH,
and Minerva for support. S.I. and R.M. thank the Horowitz complexity
science foundation PhD fellowship.

\appendix
\section{Detection of network motifs using randomized networks that preserve clustering sequences}
The set of candidate CGUs should ideally be the complete set of
subgraphs of different sizes found in the network. The complete set
of subgraphs is, however, too large for the optimization procedure
to effectively work in practice (there are 199 4-node connected
directed subgraph types, 9,364 5-node subgraph types, 1,530,843
6-node subgraph types etc., a significant fraction of which actually
occur in the real networks). Due to computational limitations, we
considered in the present study only a small subset of the
subgraphs, those which occur significantly more often in the network
than in suitably randomized networks. These subgraphs are termed
network motifs~\cite{Shen-orr,Milo,Milo 2004,Kashtan}.\\
For the detection of network motifs we considered two randomized
ensembles :(1) random networks in which each node preserves the
number of incoming, outgoing and mutual edges (edges that run in
both direction) connected to it in the real network.(2) Random
networks in which each node preserves the number of incoming,
outgoing and mutual edges connected to it in the real network, and
in addition each node preserves the clustering coefficient of that
node in the real network~\cite{Strogatz
2001}-\cite{newman_siam},\cite{Ravasz}. The detection of network
motifs, using ensemble (1) as a null hypotheses was described
in~\cite{Shen-orr,Milo,Milo 2004,Kashtan}. The random networks
created this way often have a different clustering coefficient for
each node than in the real network. As a result, the number of
nondirected triangles in the real network is generally different
from the randomized network ensemble (either higher, as in the
transistor network, or lower, as in the protein signaling network).
To control for this, in the more stringent ensemble (2), we preserve
also the clustering coefficient of each node~\cite{Strogatz
2001,Albert,newman_siam,Ravasz} ("clustering sequence") , using a
simulated annealing algorithm.
\\To create such an ensemble of randomized networks we first
randomize the real network with a Markov-chain Monte-Carlo
algorithm, which successively selects two node pairs and performs a
"switch", rewiring their edges, as described in
~\cite{Milo,milo_condmat}. To define the clustering sequence of a
network of $N$ nodes: $\{{C_i\}}_{i=1}^N$, we treat its nondirected
version~\cite{Ravasz}:
\begin{eqnarray}
C_i=\frac{2n_i}{K_i(K_i-1)}
\end{eqnarray}
$K_i$ is the number of edges connected to node $i$ (which
represent either incoming,outgoing or mutual edges in the directed
version). $n_i$ is the number of triangles connected to node $i$.
Denoting the clustering sequence of the random networks by
$\{{C_i^R\}}_{i=1}^N$ we carry out switches, again randomly
selecting pairs of edges and rewiring them, but this time with
probability :
\begin{eqnarray}
\min\{1,e^{-\Delta{E}/T}\}
\end{eqnarray}
where $T$ is an effective temperature, lowered by a factor of 5\%
between sweeps, and $E$, the energy function, is the distance
between the clustering sequences of the real and random networks :
\begin{eqnarray}
E=\sum_{i=1}^N{\frac{|C_i-C_i^R|}{C_i+C_i^R}}
\end{eqnarray}
The random networks obtained have precisely the same clustering
sequence and degree sequences as the real network. They are thus
more constrained than in ensemble (1). In the presently studied
networks, they contain almost precisely the same number of
nondirected triangles as the real network. However, the numbers of
directed triangle subtypes differ from the real network. There are 7
types of directed 3-node triangle subgraphs~\cite{Milo}. The
relative abundance of these 7 subgraphs in the random ensemble is
determined by different moments of the degree
sequences~\cite{Itzkovitz}. Thus, 3-node directed subgraphs can
still be found as motifs using ensemble (2), depending on the
network degree sequences. For the transistor network and signalling
network studied, the two sets of network-motifs of 3-6 nodes
detected using ensembles (1) and (2) had an overlap of more than
90\%. Using ensemble (2) on the transistor network results in
somewhat fewer motifs that are triangles with dangling edges, and
more tree-like motifs than ensemble (1). Using ensemble (2) on the
protein signaling network results in somewhat fewer tree-like
motifs. For both networks, the coarse-graining algorithm detected
the same optimal sets of CGUs using either ensembles. Thus, in the
present examples, coarse-graining is not affected by choice of
random network ensemble.
\section{Overlap rules}
    The desired CGU set should have minimal overlap (shared nodes)
between occurrences of the CGUs. In cases where shared nodes are
necessary, the CGU partitioning should be such that the shared nodes
do not affect the function of each CGU. The solutions that maximize
Eq. 2 or Eq. 8 often have significant overlap between the CGUs. Here
we describe rules that disqualify solutions in which overlap would
interfere with the coarse grained representation. We also describe
how an acceptable CGU partitioning is performed in
cases where overlap is allowed.\\
    Once a set of CGUs which maximizes the scoring function is found,
it is tested for the following criterion:  Allowed solutions are
those in which each overlapping node receives inputs from only one
CGU (Fig 11). CGU sets which don't meet this criterion are
disqualified, and a new set is sought. (Note that the overlapping
nodes are allowed to send outputs to both CGUs).
\begin{figure}[!hbp]
\begin{center}
\includegraphics[width = 60 mm, height = 100 mm ]{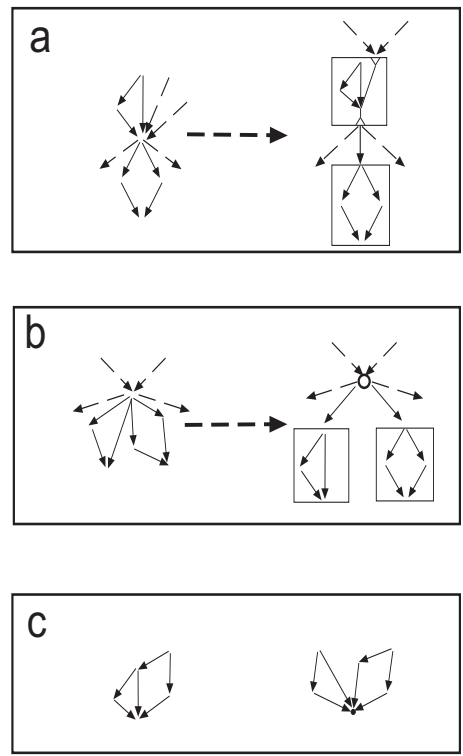}
%\resizebox{8cm}{!}{\includegraphics{Fig1.eps}}
\end{center}
\caption{Overlap rules of CGU candidates. In these examples the CGU
candidates are \textbf{a.} A 3-node feed-forward loop (left) and a
4-node diamond subgraph (right). \textbf{b.} Overlap of nodes which
receive inputs from only one of the CGUs(left), and coarse-grained
representation (right). \textbf{c.} Overlap of nodes which send
outputs to two CGUs (left), and coarse-grained representation
(right). Note the addition of a node upstream of the two CGUs,
marked with an open circle. \textbf{d.} Two examples of disqualified
cases, were a node receives inputs from both CGUs : two CGUs with a
common edge (left), and without a common edge (right). }
\label{fig11}
\end{figure}
    In acceptable CGU sets, in every case of an overlap,
the overlapping node is duplicated, and appears once in each of
the CGU occurrences. The acceptance criterion above ensures that
the inputs to the duplicated nodes can be fully captured by one of
the CGUs, thus ensuring that the function of the coarse grained
network can be inferred from the functions of individual CGUs.
Finally, in cases where the overlapping node only sends output to
the CGUs and does not receive inputs from them, an additional node
is created in the coarse-grained network. This node has all of the
connections of the original node, and sends outputs to the
duplicated node in each CGU (e.g. MKP2 and MKP5 in Fig9,10 and
fig11c)
\section{Algorithm for detecting PGNMs}
Topological generalizations are subgraphs obtained from smaller
network motifs, by replicating one or more of their roles, together
with their connections~\cite{Kashtan}. The role-replication
operation does not increase the number of roles in the resulting
generalized subgraph, which maintains a perfect fit to the block
model of the network motif. Additionally each node has the same role
in both the generalized subgraph and in every occurrence of the
basic motif included in it. The role assignment is thus
automatically defined. Probabilistically generalized network motifs
(PGNM's) are subgraphs
which have a small distance $d$ (Eq. 5) from its block model.\\
To detect PGNMs we start with a network motif $\mu$. The nodes of
each occurrence of $\mu$ are partitioned into roles \cite{Kashtan}.
We then form a nondirected graph $R_{\mu}$ in which each node,
$r_{\mu}^i$ is an occurrence of $\mu$ in the original network $R$,
and a nondirected edge between two nodes $r_{\mu}^i$ and $r_{\mu}^j$
is set if a) any of the nodes of these occurrences in the original
network $R$ are connected by an edge, or b) any of the nodes in the
original network overlap. After establishing $R_{\mu}$ we start from
each node $r_{\mu}^i$ and perform a search, consecutively adding the
one node in $R_{\mu}$ which provides the best fit to the block model
of $\mu$ (the resulting joined subgraph with the smallest increase
in $d$). We stop when $d$ is greater than a threshold (we use
$0.3$). When calculating the fit to the block-model, we partition
the nodes of the joined subgraph according to their role assignment
in $\mu$. If a node in $R$ has different roles in two different
occurrences of $\mu$, when calculating $d$ for the joined subgraph,
we take the smallest distance obtained from all possible labellings
of this node (for example, nodes $3,4$ in subgraph $G2$ of Fig $6$
share role $1$ in the bifan $(3,4,5,6)$ and role $2$ in the bifan
$(1,2,3,4)$). We iterate this procedure by beginning with each
$r_{\mu}$, establishing a list of embedded subgraphs (if two
embedded structures have the same $d$ we keep only the larger one).
These subgraphs are probabilistic generalizations of $\mu$, tagged
by their distance from a perfect generalization, $d$. In finding the
optimal coarse-graining we perform a simulated annealing algorithm,
sequentially generating a new active set of CGUs, recalculating the
scoring function (Eq. 8) and accepting the new active set with a
Metropolis Monte-Carlo probability. During the optimization, we also
test the  resulting score from coarse-graining only subsets of the
PGNMs
of each CGU. For an alternative definition of probabilistic network motifs see \cite{berg}\\


\begin{thebibliography}{99}
\bibitem{footnote0}
Two subgraph occurrences with connectivity profile vectors $V_1$
and $V_2$ are grouped together if there exists a permutation $P$
of the nodes that preserves the subgraph structure and for which
the permuted profile vectors obey $P(V_1)=V_2$.
\bibitem{footnote1}
The new active set is accepted with probability
$\min\{1,e^{\Delta{S}/T}\}$, where $\Delta{S}$ is the score
difference from the previous active set, and $T$ is an effective
 temperature, lowered by a factor of 5\% between sweeps.
\bibitem{footnote2}
This distance measure accounts for the size of the subgraph, and is
more appropriate than measures such as the Hamming distance (number
of edges which have to be added or removed to obtain a perfect fit
to a block-model).
\bibitem{footnote3}
When establishing the CGU candidate set we also ignore the
connectivity profile of subgraphs. This further softens the CGU
criteria. This scoring function still favors units with few mixed
nodes, because they affect the number of ports, $P$ in the
coarse-grained network.
\end{thebibliography}
\end{document}